\def \be {\begin{equation}}
\def \ee {\end{equation}}
\def \bea {\begin{eqnarray}}
\def \eea {\end{eqnarray}}
\def \nn {\nonumber}

\def \la {\langle}
\def \ra {\rangle}
\def \R {{\bf R}}
\def \C {{\bf C}}
\def \Z {{\bf Z}}
\def \del {\partial}
\def \dels {\partial\kern-.5em / \kern.5em}
\def \As {{A\kern-.5em / \kern.5em}}
\def \Ds {D\kern-.7em / \kern.5em}

\def \a {\alpha}
\def \b {\beta}
\def \dag {\dagger}
\def \g {\gamma}

\def \d {\delta}
\def \eps {\epsilon}
\def \Eps {{\cal E}}

\def \s {\sigma}

\def \th {\theta}

\def \II {I\hspace{-.1em}I\hspace{.1em}}

\def \IIB {\mbox{\II B\hspace{.2em}}}

\def \zb {\bar{z}}
\def \H {{\cal H}}
\def \A {{\cal A}}
\def \D {{\cal D}}
\def \xh {\hat{x}}
\def \yh {\hat{y}}
\def \zh {\hat{z}}
\def \zbh {\hat{\zb}}
\def \fh {\hat{f}}
\def \gh {\hat{g}}
\def \nh {\hat{n}}
\def \delh {\hat{\del}}

\def \delb {\bar{\del}}

\def \SHO {\mbox{SHO}}
\def \GNS {\mbox{GNS}}

\documentclass[12pt]{article}

\begin{document}
\begin{titlepage}

\begin{center}
\hfill hep-th/0003012\\
\vskip .5in

\textbf{\Large Twisted Bundle on Noncommutative Space \\
               and $U(1)$ Instanton}

\vskip .5in
{\large Pei-Ming Ho}
\vskip 15pt

{\small \em Department of Physics,\\
National Taiwan University,\\
Taipei 106,\\
Taiwan, R.O.C.}

\vskip .2in
\sffamily{pmho@phys.ntu.edu.tw}

\vspace{60pt}
\end{center}
\begin{abstract}

We study the notion of twisted bundles
on noncommutative space.
Due to the existence of projective operators in
the algebra of functions on the noncommutative space,
there are twisted bundles with non-constant dimension.
The $U(1)$ instanton solution of Nekrasov and Schwarz
is such an example.
As a mathematical motivation for not excluding such bundles,
we find gauge transformations by which
a bundle with constant dimension can be equivalent to
a bundle with non-constant dimension.

\end{abstract}
\end{titlepage}
\setcounter{footnote}{0}

\section{Introduction}

Noncommutative geometry \cite{Con}
has been applied to string theory in several different aspects
\cite{Wit}-\cite{CHL}.
In particular, Connes, Douglas and Schwarz \cite{CDS} found that
in the matrix theory a constant $C_{-ij}$ field background
results in a gauge field theory living on a noncommutative space.
It was also found that the low energy effective theory
for D-branes in the NS-NS $B$ field background is also
given by the same noncommutative gauge theory (NCGT)
\cite{AAS,CH1,Schom,SW}.
We feel that it is important to understand
the physics of noncommutative field theories.

The purpose of this paper is to understand better
twisted bundles on noncommutative space.
A noncommutative space is defined by
the algebra $\A$ of functions on the space,
and the Hilbert space on which $\A$ is realized as operators.
According to Connes,
vector bundles on a noncommutative space
are defined as projective modules of $\A$ \cite{Con}.
In other words, a twisted bundle can be understood
as a projection $P$ from a trivial bundle.
Classically we restrict ourselves to bundles
with fibers all of the same dimension at every point.
This means that the projection $P$ is of constant rank.
However, due to the existence of projective operators in $\A$,
one can extend the definition of bundles
to projective modules for which $P$ has non-constant rank.
The $U(1)$ instanton solution of Nekrasov and Schwarz \cite{NS}
is found to be such an example \cite{Fur}.

In sec.\ref{Hilbert} we review the noncommutative algebra $\A$
of functions on a quantum plane,
and its realization on Hilbert spaces.
In sec.\ref{algebra}, we find projective operators in $\A$
constructed out of operators $U$
which has right inverse but no left inverse.
Formal gauge transformations by $U$ bring
a bundle of constant dimension to a bundle with non-constant dimension.
This can be viewed as a mathematical motivation for us
not to exclude bundles with non-constant dimension.
We review the notion of twisted bundles in sec.\ref{proj-mod},
with special care to projective modules
with non-constant rank \cite{Fur}.
As an example we review the ADHM construction of
$U(1)$ instantons \cite{NS} in sec.\ref{ADHM}.

\section{Quantum Plane} \label{Hilbert}

In Connes' formulation of noncommutative geometry \cite{Con},
a noncommutative space is specified by the spectral triple $(\A,\H,\D)$,
where $\A$ is the noncommutative algebra of functions on the space,
$\H$ is the Hilbert space on which $\A$ is realized as an operator algebra,
and $\D$ is the so-called Dirac operator which defines
the metric and differential calculus.
\footnote{
For a short introduction to the role $\D$ plays in noncommutative geometry
please see \cite{HW1}.
}
In this paper, we will not use $\D$,
but instead we will directly define the differential calculus
for the nonocommutative space under consideration.

For the noncommutative space we are interested in,
the coordinates $x^i$ satisfy
\be \label{NC}
[x_i, x_j]=i\theta_{ij},
\ee
where $\th_{ij}\in\R$ are constant real numbers.
The algebra $\A$ is defined by this relation
and elements in $\A$ are functions of $x_i$.

There are two different Hilbert spaces
that are often used in the literature for our $\A$.
The first is the Hilbert space of simple harmonic oscillators (SHO),
on which complex coordinates act as creation and annihilation operators.
The second is the phase space of one particle on a classical space,
on which the noncommutative coordinates are linear combinations of
coordinates and momenta in quantum mechanics.
We discuss these Hilbert spaces in the following two subsection.

For simplicity, we will consider a two dimensional noncommutative space,
which is also called a quantum plane,
with coordinates $\xh$ and $\yh$ satisfying
\be
[\xh,\yh]=i\th.
\ee
This relation is invariant under translation and rotation
on the two dimensional plane.
There is an anti-involution denoted by $\dag$
which is identified with Hermitian conjugation when
$\A$ is realized as operators on a Hilbert space.
We assume that $\th$ is a real number and
\be
\xh^{\dag}=\xh, \quad \yh^{\dag}=\yh.
\ee
The algebra $\A$ is generated by $\xh$ and $\yh$.
As a convention we choose the coordinates such that
\be
\th \geq 0.
\ee
It follows that the commutation relation of the complex coordinates
\be
\zh=\frac{\xh+i\yh}{2}, \quad \zbh\equiv\zh^{\dag}=\frac{\xh-i\yh}{2}
\ee
is similar to that of the creation and annihilation operators
for a simple harmonic oscillator:
\be \label{zzb}
[\zh,\zbh]=\frac{\th}{2}.
\ee
One can therefore use the SHO Hilbert space as a representation of $\A$.

One can adjoin differential one-forms $(d\zh)$ and $(d\zbh)$
to this algebra and generate the differential calculus.
By definition, $(d\zh)$ and $(d\zbh)$ commute with $\zh$ and $\zbh$,
and satisfy the anticommuting relations
\be
(d\zh)(d\zbh)=-(d\zbh)(d\zh), \quad (d\zh)^2=(d\zbh)^2=0.
\ee
One can also define the derivatives $\delh_i$ ($i=1,2$) which satisfy
\bea
&\delh_i \xh_j = \xh_j \delh_i + \d_{ij}, \\
&[\delh_1, \delh_2] = 0.
\eea
The action of a derivative on a function is
\be
(\delh_i f)\equiv [\delh_i, f].
\ee
When acting on functions, $\delh_i\sim\th^{-1}_{ij}\xh_j$.

\subsection{SHO Hilbert Space} \label{SHOH}

The algebra $\A$ can be realized as an operator algebra according to
\be \label{SHO}
\zh|n\ra=\sqrt{\frac{\th n}{2}}|n-1\ra, \quad
\zbh|n\ra=\sqrt{\frac{\th(n+1)}{2}}|n+1\ra
\ee
on the SHO Hilbert space $\H_{\SHO}=\{ |n\ra, n=0,1,2,\cdots \}$.

This realization of $\A$ is not a one-to-one map.
Any smooth function of
\be
\nh=\frac{2}{\th}\zbh\zh,
\ee
such as $\sin{\pi\nh}$,
which vanishes for all $\nh\in\Z$ is identical to zero.
Similarly, a smooth function of $\nh$
which vanishes at all integers except $\nh=n$
is equivalent to the projective operator $P_n=|n\ra\la n|$.

Intuitively, points on the noncommutative space
correspond to states in the Hilbert space.
The existence of these projective operators
suggests that in some sense
the noncommutative space is a disconnected space.
Classically, the step function is a projective operator,
but it is not a smooth function.

In general, a necessary condition for a Hilbert space $\H$
to respect a symmetry $\xh\rightarrow \xh'=\xh'(\xh)$ is that
\be
Tr(f(\xh'))=Tr(f(\xh))
\ee
for the trace on $\H$.
Note that the natural notion for intergration on a quantum space is
just the trace over its Hilbert space.
\footnote{To be more precise one should use the Dixmier trace.}

Denoting the trace of $f\in\A$ on a Hilbert space
for the quantum plane by $\la f(\xh,\yh)\ra$,
one can check whether it is true that
\be \label{transl-inv}
\la f(\xh+a,\yh+b)\ra=\la f(\xh,\yh)\ra
\ee
for arbitrary real constants $a, b$
whenever $\la f\ra$ is well defined.
This requirement is equivalent to
\be \label{tr-inv2}
\la(\delh_i f)\ra = 0,
\ee
which is the quantum analogue of the Stokes' theorem.
It can be checked that
the trace over $\H_{\SHO}$
respects both the rotational 
and the translational symmetry.

\subsection{GNS Construction}

The problem here is generic.
Given a noncommutative algebra of functions $\A$,
how do we define its Hilbert space such that
certain symmetries of the algebra are respected?
The natural solution is the following.
First define the integration $\la\cdot\ra$ on the noncommutative space
as a functional invariant under all required symmetries,
then use the Gel'fand-Na\v{i}mark-Siegel (GNS) construction
to build the Hilbert space from this functional.
In the GNS construction, the Hilbert space $\H_{\GNS}$ is
roughly speaking the set of all functions $\{ |f(\xh)\ra \}$,
where the inner product is defined by
\be
\la f(\xh)|g(\xh)\ra = \la f(\xh)^{\dag}g(\xh)\ra.
\ee
However, only normalizable states should be considered,
and two states are viewed equivalent
if they have the same inner product with all other states.


Let $\xh_1=\xh, \xh_2=\yh$.
The operators $\xh_i$ can be realized
in terms of the coordinates $x_i$ and derivatives $\del_i$
on a classical plane as
\be \label{xh-x}
\xh_i=x_i+\frac{i}{2}\th_{ij}\del_j,
\ee
where $\th_{ij}=\th\eps_{ij}$.
Define the ``vacuum'' $\ra$
which is annihilated by the derivatives \cite{Zum,CHL}
\be \label{vac}
\del_i\ra=0.
\ee
The vacuum correpsonds to the identity function $1$
and a classical function $f$ acts on it to give a new state $f(x)\ra$.
The vacuum may not be a state in $\H$ and is simply a covenient notation.

For any function $f$ of $\xh$,
there is a classical function $f_c$ of $x$ such that
\be \label{fc}
f(\xh)\ra = f_c(x)\ra
\ee
due to (\ref{vac}).
The function $f_c$ is also the classical function corresponding
to $f(\xh)$ in the star product representation (see the next subsection).
We can now define the integration $\la f(\xh)\ra$ on
the quantum plane by the classical integration of $f_c(x)$.
Since according to (\ref{xh-x}),
the translation of $\xh$ is identical to the translation of $x$,
this functional $\la\cdot\ra$ preserves translational symmetry.

This integral is different from the integral on $\H_{\SHO}$.
For instance, any function of $\zbh\zh$ which vanishes
when $(\th\zbh\zh/2)$ is integer is always integrated to zero on $\H_{\SHO}$,
but not necessarily zero on $\H_{\GNS}$.

Following the GNS construction with this invariant functional,
we obtain the Hilbert space $\H_{\GNS}$
which is the same as the Hilbert space of
one-particle quantum mechanics on a classical space.

$\H_{\GNS}$ is not an irreducible representation of $\A$,
and $\H_{\SHO}$ is a subset of this Hilbert space
corresponding to picking up only the states
$\{ |n\ra\equiv H_n(x/\sqrt{\th})e^{-x^2/\th}\ra \}$
where the $H_n$'s are the Hermite polynomials.

Due to (\ref{vac}), the requirement (\ref{tr-inv2})
is consistent with viewing the functional
as the ``vacuum expectation value'',
with the assumption that
\be
\la\delh_i=0.
\ee

It seems that
$\H_{\GNS}$ is the correct choice
to be used in matrix theory with $C$ field background
or for D-branes in a $B$ field background.
In the derivation of the noncommutative relations (\ref{NC})
via open string quantization \cite{CH1,CH2},
the noncommutativity comes from a mixing
of classical coordinates and momenta due to
mixed boundary conditions for the open string.
This is very much like what eq.(\ref{xh-x}) represents.

\subsection{Star Product Representation}

It is well known that the quantum plane can be described by
commutative functions with a star product.
Let $x$ and $y$ be the coordinates over a classical plane.
The star product is defined by
\be
f\ast g = \cdot
\left(e^{i\frac{\th}{2}(\del_x\otimes\del_y-\del_y\otimes\del_x)}
(f\otimes g)\right),
\ee
where $\cdot(A\otimes B)$ is
the ordinary commutative product of $A$ and $B$.

For each classical function $f(x)$
we associate a pseudo-differential operator \cite{CHL}
\be
\fh = \sum_{n=0}^{\infty}i^{n}\frac{\th_{i_1 j_1}}{2}\cdots
\frac{\th_{i_n j_n}}{2}\left(\del_{i_1}\cdots\del_{i_n}f(x)\right)
\del_{j_1}\cdots\del_{j_n},
\ee
which can be written as
\be
\fh = \cdot
\left(e^{i\frac{\th}{2}(\del_x\otimes\del_y-\del_y\otimes\del_x)}
(f\otimes\cdot)\right).
\ee
It can be shown that in general
\be
\fh \gh = \hat{h}, \quad \mbox{where}\;\; h=f\ast g.
\ee

Let the Fourier transform of $f$ be denoted $\tilde{f}_k$:
\be
f(x)=\int dk \tilde{f}_k e^{ik\cdot x},
\ee
then one finds
\be \label{fk}
\fh = \int dk \tilde{f}_k e^{ik\cdot\xh},
\ee
where $\xh$ is defined in (\ref{xh-x}).
Thus $\fh$ is a function of $\xh$,
and the algebra of classical functions with the star product
is equivalent to $\A$.
It is straightforward to check from (\ref{fk}) that
\be
\fh(\xh)\ra = f(x)\ra.
\ee
Comparing this with (\ref{fc}), one sees that $f=f_c$.

\section{An Algebraic Problem} \label{algebra}

Consider the standard algebraic problem
of finding all solutions of $\psi$ from an equation like
\be \label{fpsi0}
f\ast\psi = 0
\ee
for given $f(x)$.
Given a special example of (\ref{fpsi0}),
one can obtain other examples by replacing
$f$ by $f'\ast f$ and $\psi$ by $\psi\ast g$
with arbitrary functions $f'$ and $g$.

In noncommutative algebra, an element $f$ may not have
a left inverse $f'$ such that $f'\ast f=1$,
or a right inverse $f'$ such that $f\ast f'=1$.
If $f$ has a left inverse, then the unique solution of $\psi$ is zero.
Conversely, if $\psi$ has a trivial solution,
it means that $f$ can not have a left inverse.

Because the star product involves derivatives,
the equation (\ref{fpsi0}) is not really an algebraic relation
but rather a differential equation.
Differential equations are uniquely solved only if
sufficient initial or boundary conditions are specified,
so we should expect that there are
generically more than one solution to the equation (\ref{fpsi0}).

Consider the example of $f=z$:
\be
z\ast\psi=z\psi+\frac{\th}{4}\delb\psi=0.
\ee
It can be easily solved by
\be
\psi=e^{-\frac{4z\zb}{\th}+h(z)},
\ee
where $h(z)$ is an arbitrary function of $z$.
The special case of $h$ equal to a constant gives
the Gaussian distribution
\be \label{Gauss}
\psi=\psi_0 e^{\pm\frac{x^2+y^2}{\th}}.
\ee
This function corresponds to the operator $|0\ra\la 0|$ on $\H_{\SHO}$.

There are also examples of $f$ for which the solution of $\psi$
in (\ref{fpsi0}) is unique ($\psi=0$).
An example is given by $f=e^{ikx}$.

It is easy to see that if $f$ is a polynomial of degree $k$
then (\ref{fpsi0}) is a differential equation of order $k$,
and one needs the appropriate initial or boundary condition
for the given differential equation.

In terms of the operator algebra on $\H_{\SHO}$,
the fact that (\ref{fpsi0}) has many solutions for $f=z^k$ 
is related to the fact that $\zh^k$ annihilates
all states $|n\ra$ with $n<k$.
Instead of using $\zh$ in this discussion,
let us normalize it and define
\be \label{u}
u=(\zbh\zh+\th/2)^{-1/2}\zh.
\ee
One can check that
\be \label{uudag}
uu^{\dag}=1, \quad \mbox{but}\;\;
u^{\dag}u=1-|0\ra\la 0|.
\ee
Thus
\be \label{Puu}
p=u^{\dag}u
\ee
is a projective operator on $\H_{\SHO}$, i.e., $pp=p$.
This is because $u$ has the annihilation operator $\zh$
on the right to annihilate $|0\ra$.
Thus $u$ has a right inverse $u^{\dag}$ and
one can prove that it has no left inverse.
The existence of this kind of operators $u$ is not
a special feature of the SHO Hilbert space.
The operator $u$ has the same properties on $\H_{\GNS}$.
In general, $(u^{\dag})^k u^k$ for positive integer $k$
is a projective operator which annihilates all $|n\ra$ with $n<k$.

In general, we are interested in operators $U$ which satisfy
\be
UU^{\dag}=1, \quad \mbox{but}\;\; U^{\dag}U\neq 1.
\ee
As above, $U^{\dag}U$ is always a projective operator.
Since $U$ does not have a left inverse,
a formal gauge transformation of the gauge potential $A$ by $U$
\be
A\rightarrow A'=UAU^{\dag}+U(dU^{\dag})
\ee
is not really a gauge transformation.
We can not transform $A'$ back to $A$ by any transformation.
It follows that a transformation by $U^{\dag}$ also
should not be viewed as a true gauge transformation.
However we will show in sec.\ref{further} that
$U$ still defines an equivalence relation
by modifying the bundle at the same time.

\section{Projective Module as Twisted Bundle} \label{proj-mod}

In noncommutative geometry, vector bundles are defined as
projective modules of the algebra $\A$ of functions on the base space.
In this section we review the notion of twisted bundles
as projective modules and derive some formulas to be used latter
for the ADHM construction.

For a trivial vector bundle of $N$ dimensional fibers,
a section of the bundle is a vector of $N$ entries in $\A$.
The set of sections is a free module $\A^{\otimes N}$.

To motivate the definition of twisted bundles,
we recall the example of the tangent bundle on $S^2$.
This tangent bundle is nontrivial,
but the direct sum of the tangent space and the normal vector space
constitute a trivial bundle of three dimensions.
One can choose a projective operator acting on this trivial bundle
which is the $3\times 3$ matrix
that projects any 3-vector on a fiber of the trivial bundle
to the tangent space.
Explicitly, the projection is given by
$P=1-|n\ra\la n|$, where $|n\ra$ is the normal vector on $S^2$.
Thus sections of the tangent bundle on $S^2$
form a projective module,
since the definition of a projective module is just that
it is a direct summand of a free module $\A^{\otimes N}$ with finite $N$.

Classically all bundles are projective modules and vise versa
due to a theorem by Serre and Swan.
\footnote{
The theorem of Serre and Swan states that all locally trivial
finite-dimensional complex vector bundles over a compact space $X$
are one-to-one corresponding to all finite projective modules
over the algebra $\A=C(X)$.
}
The natural definition of bundles on noncommutative spaces
is thus just projective modules \cite{Con}.
Given a bundle, one can always find an $N \times N$ matrix $P$
with elements in $\A$ which satisfies $PP=P$,
such that the set of sections on the bundle is
the projective module $P\A^{\otimes N}$ for some integer $N$.
\footnote{
$P\A^{\otimes N}$ is a projective module because
its direct sum with $(1-P)\A^{\otimes N}$ is a free module.
}
An obvious property of projective modules is that
since the projection is taken from the left by convention,
it is a right $\A$ module because
one can multiply elements in $\A$ from the right,
to each copy of $\A$ in $\A^{\otimes N}$.

In practice it is not always necessary to define a bundle
in terms of the projection $P$ and free module $\A^{\otimes N}$.
For instance the $U(N)$ twisted bundles on a noncommutative torus
are constructed in \cite{Ho} directly without referring to $P$ or
the trivial bundle.

\subsection{Connection From Projection}

Any connection on the trivial bundle induces
a connection on the twisted bundle via this projection.
A familiar example is how the tangent bundle of a curved space 
gets its connection from its embedding in a flat space.
We will give explicit expressions for the covariant derivative
on the twisted bundle,
provided that the projection and free module are given.

A section on a trivial bundle $\Eps_0$ of dimension $N_0$
can be denoted as $|s(x)\ra = |a\ra s_a(x)$,
$a=1,2,\cdots,N_0$,
where $s_a(x)\in\A$ are functions on the base space.
A section is an element of the free module $\A^{\otimes N_0}$.
Let the covariant derivative on $\Eps_0$ be $D=d+A$,
where $A$ is an $N_0\times N_0$ matrix of elements in $\A$.
For simplicity, we first assume that the basis $\{ |a\ra \}$
is orthonormal and covariantly constant.
That is, it satisfies
\bea
&\la a|b\ra = \d_{ab}, \quad a,b=1,2,\cdots,N_0, \\
&(D|a\ra) \equiv (d|a\ra)+|b\ra A_{ba}(x) = 0, \label{Da}
\eea
where $A_{ba}=\la b|A|a\ra$.
The action of $D$ should satisfy the Leibniz rule
\be
(D|s\ra f)=(D|s\ra)f+|s\ra (df)
\ee
for any section $|s\ra$ and function $f$.
It follows from (\ref{Da}) that
\be
(D|s(x)\ra) = |a\ra (ds_a(x)).
\ee

Consider an arbitrary projection $P$
which can be ``locally''
\footnote{
What is meant by ``locally'' is not obvious
for noncommutative spaces in general.
A special technique is developed in \cite{CHZ}.
For the case of instantons it is clear that
by this we mean the quantum $\R^4$,
which is a local patch of $S^4$.
}
expressed as
\be \label{P}
P=\sum_{i=1}^{N} |\psi_i(x)\ra\la \psi_i(x)|,
\ee
where $N$ is the rank of the projector ($N\leq N_0$),
and
\be
|\psi_i(x)\ra=|a\ra \psi_{ai}(x)
\ee
are sections of $\Eps_0$.
Until the next subsection we assume that
\be \label{ortho}
\la\psi_i|\psi_j\ra=\sum_a \psi^{\dag}_{ai}(x)\psi_{aj}(x) = \d_{ij}
\ee
so that
\be
PP=P.
\ee

The projection $P$ defines a projective module $P\A^{\otimes N_0}$
which corresponds to a new bundle $\Eps$.
Sections on $\Eps$ are of the form
\be \label{sec}
|u(x)\ra = |\psi_i\ra u_i(x)
\ee
for $u_i(x)\in\A$ ($i=1,2,\cdots,N$).

The induced covariant derivative on $\Eps$ is
\be
D_P=PD.
\ee
When acting on $\Eps$,
\bea
(D_P|u(x)\ra) &=& (PD|\psi_i\ra u_i) \nn \\
              &=& (PD|a\ra\psi_{ai} u_i) \nn \\
              &=& P|a\ra(d\psi_{ai} u_i) \nn \\
              &=& |\psi_j\ra\psi_{aj}^{\dag}(d\psi_{ai} u_i) \nn \\
              &=& |\psi_i\ra (D_P u)_i,
\eea
where
\bea
(D_P u)_i(x) &=& (d u_i(x)) + (A_P)_{ij} u_j(x), \\
(A_P)_{ij} &=& \psi_{ai}^{\dag}(x)(d\psi_{aj}(x))
           = \la\psi_i|d|\psi_j\ra.  \label{AP}
\eea
This is the connection on a twisted bundle
obtained as a projection from a free module.
Eq.(\ref{AP}) is reminisent of the Berry phase.

If the orthonormal basis $\{ |a\ra \}$ does not satisfy (\ref{Da}),
but rather $D|a\ra = |b\ra A'_{ab}$,
then (\ref{AP}) will be modified by an additional term
$A'_{ij} u_j$ where
$A'_{ij}=\la i|A'|j\ra=\psi_{ai}^{\dag}A'_{ab}\psi_{bj}$.

The field strength is by definition
\be
F=(dA) + AA
\ee
for a given gauge potential $A$. 
Using (\ref{P}) and (\ref{ortho}),
the field strength of (\ref{AP}) is
\bea \label{FP1}
(F_P)_{ij} &=& (d\psi_{ai}^{\dag})(d\psi_{aj})
   +\psi_{ai}^{\dag}(d\psi_{ak})\psi_{bk}^{\dag}(d\psi_{bj}) \nn\\
           &=& (d\psi_{ai}^{\dag})(d\psi_{aj})
   -(d\psi_{ai}^{\dag})\psi_{ak}\psi_{bk}^{\dag}(d\psi_{bj}) \nn\\
           &=& (d\psi_{ai}^{\dag})(\d_{ab}-P_{ab})(d\psi_{bj}),
\eea
where $P_{ab}=\la a|P|b\ra=\psi_{ak}\psi^{\dag}_{bk}$.

The projective operator $(1 - P)$ in (\ref{FP1})
can be written in terms of a basis of vectors as
\be \label{one-P}
(1-P)=\sum_{\a=1}^{N_0-N}|\phi_{\a}\ra\la\phi_{\a}|,
\ee
where $|\phi_{\a}\ra=|a\ra\phi_{a\a}(x)$ satisfy
\bea \label{phipsi}
&\la\phi_{\a}|\psi_i\ra = 0, \\
&\la\phi_{\a}|\phi_{\b}\ra = \d_{\a\b}, \label{phiphi}
\eea
for all $\a, \b=1,2,\cdots,(N_0-N)$, $i=1,2,\cdots,N$.
So $\{ |\psi_i\ra\}$ is a basis for $\Eps$,
and together with $\{ |\phi_{\a}\ra\}$ they give
a complete basis for $\Eps_0$.
Using (\ref{one-P}) and (\ref{phipsi}),
the field strength (\ref{FP1}) can be written as
\be \label{FP2}
(F_P)_{ij}=\psi_{ai}^{\dag}(d\phi_{a\a})(d\phi_{b\a}^{\dag})\psi_{bj}.
\ee

In the ADHM construction of instanton solutions,
we are given conditions of the form
\be \label{xipsi}
\xi^{\dag}_{a\a}\psi_{a} = 0,
\ee
where $\xi_{\a}$ may not be orthonormal,
but can be related to the orthonormal basis $\phi_{\a}$
by a linear transformation
\be
\xi_{a\a}= \phi_{a\b} M_{\b\a},
\ee
where $M$ satisfies
\be \label{MM}
\xi^{\dag}_{a\a}\xi_{a\b} = M_{\a\g}^{\dag}M_{\g\b}
\ee
as a result of (\ref{phiphi}).
The solutions of $\psi$ for (\ref{xipsi}) are labelled as $\psi_i$,
which defines a projective operator as (\ref{P}).
Thus eq.(\ref{xipsi}) can be viewed as
defining equations for a twisted bundle.
Using (\ref{xipsi}) and (\ref{MM}),
we derive the expression for $F$ as
\be \label{FP3}
(F_P)_{ij}=\psi_{ai}^{\dag}
(d\xi_{a\a})(\xi^{\dag}\xi)^{-1}_{\a\b}(d\xi_{b\b}^{\dag})
\psi_{bj}.
\ee
The expression (\ref{FP3}) is particularly useful
for the ADHM construction of instanton solutions.

The calculations above are valid even for noncommutative algebras
because we have only assumed associativity of the algebra $\A$.

\subsection{Projections of Non-Constant Rank} \label{further}

There are two kinds of projective operators depending on
whether they have a constant rank through out the space.
For instance, for the free module $\A$ ($N=1$)
on a classical space which has disconnected pieces,
one can define a projection
which equals $1$ on one piece
and equals $0$ on another.
This projection is a smooth function in $\A$,
so it defines a projective module.
However, the corresponding bundle does not have
a constant dimension everywhere on the space.
It is of one dimension on one piece
and zero dimension (no bundle at all) on another.
As we have seen in sec.\ref{algebra},
for a noncommutative space there can be elements in $\A$
which are projective even though the classical limit of
this space is simply connected.
The question is: whether projective modules associated
with these projections should be forbidden,
or how different they are from others.
In this subsection we first give a description
of bundles with non-constant dimension,
and then show that some of these bundles may be related
to bundles with constant dimension via gauge transformations.
This suggests that one should not exclude these bundles
in noncommutative gauge theories.
The appropriate formulation for these bundles
is given by Furuuchi \cite{Fur},
where it is also noted that the $U(1)$ instanton solution
discovered by Nekrasov and Schwarz in \cite{NS} is of this kind.
We will have more discussions on it in sec.\ref{instanton}.

In the previous subsection,
we have assumed that the rank of the projection $P$
is constant in (\ref{ortho}).
To be more precise about what we mean,
recall that a state in the Hilbert space $\H$
can be roughly speaking viewed as a fuzzy point
on the noncommutative space.
One can evaluate a function in $\A$ at a fuzzy point
by taking the expectation value of it for the corresponding state.
If (\ref{ortho}) is satisfied,
$|\psi_i\ra$ is nonzero everywhere,
and the rank of $P$ is $N$ everywhere.

To include the projections with non-constant rank,
we have to replace the condition (\ref{ortho})
by a weaker one in the derivation above.
To simplify our notation,
we will omit all indices on matrices.
In the following $\psi$ represents an $N_0\times N$ matrix,
and $\phi$ represents an $N_0\times N'$ matrix.
$N'$ is not smaller than $(N_0-N)$.
Assume that we are given $\phi$ and $\psi$ satisfying
\bea
&\phi\phi^{\dag}+\psi\psi^{\dag}=1, \label{compl} \\
&\phi^{\dag}\psi=0, \label{constr} \\
&\psi^{\dag}\psi\psi^{\dag}=\psi^{\dag}, \quad
 \psi\psi^{\dag}\psi=\psi. \label{proj}
\eea
The first two conditions are the same as
(\ref{one-P}) and (\ref{phipsi}) which were used before.
The condition (\ref{proj}) insures that
\be
p\equiv \psi^{\dag}\psi
\ee
is projective, i.e. $pp=p$.
It is weaker than (\ref{ortho}).

Now consider the bundle $\Eps$ composed of sections
$|u\ra=|\psi_i\ra u_i$ with $u_i\in\A$ satisfying
\be \label{pu}
pu_i=u_i.
\ee
Let $D=d+A$ with
\be
A=\psi^{\dag}(d\psi)
\ee
as in (\ref{AP}),
and define the covariant derivative on $\Eps$ by
\bea
D_p&\equiv&pDp \nn\\
   &=&\psi^{\dag}d\psi, \label{Dp}
\eea
where we have used (\ref{proj}).
In the above we have used the notation that
the action of the exterior derivative is denoted by parentheses,
so that $df=(df)+fd$ for a function or even differential form $f$,
and $dA=(dA)-Ad$ for an odd differential form $A$.
When acting on sections of the bundle,
\be
(D_p|u\ra)=(du)+A_p u,
\ee
where
\be
A_p=-(d\psi^{\dag})\psi.
\ee
Note that we have to define the bundle by (\ref{pu})
in order for the action of $D_p$ to be decomposed as $(d+A_p)$.
This in general will not be possible
if one defines the bundle by all sections of the form (\ref{sec}).
By imposing (\ref{pu}), the bundle has zero dimension at
states which is projected out by $p$,
so the bundle has non-constant dimension.

The field strength can be calculated as
\bea
F_p&\equiv&D_p^2 \nn \\
   &=&\psi^{\dag}d\psi\psi^{\dag}d\psi \nn \\
   &=&\psi^{\dag}d\psi\psi^{\dag}((d\psi)+\psi d) \nn \\
   &=&\psi^{\dag}((d\psi\psi^{\dag}(d\psi))-(d\psi)d)+d\psi d \nn \\
   &=&\psi^{\dag}((d\psi\psi^{\dag})(d\psi)+\psi dd) \nn \\
   &=&((d\psi^{\dag})-(d\psi^{\dag})\psi\psi^{\dag})(d\psi) \nn \\
   &=&\psi^{\dag}(d\phi)(d\phi^{\dag})\psi, \label{Fp}
\eea
where we used (\ref{compl})-(\ref{proj}) and $dd=0$.
This expression for the field strength is exactly the same as (\ref{FP2}).
It follows that (\ref{FP3}) is also valid when
a linear combination of $\phi$ is used.

Suppose we are given $\phi$ and $\psi$
satisfying (\ref{compl})-(\ref{proj}),
then we can find new sets of $\phi$ and $\psi$
by the formal gauge transformation
\be
\psi\rightarrow \psi U, \quad \phi\rightarrow \phi,
\ee
where $U$ is an $N\times N$ matrix of elements in $\A$ satisfying
\be
UU^{\dag}=1.
\ee
It follows that all three conditions are still valid,
and so by (\ref{Dp}) and (\ref{Fp})
\be \label{uFu}
D_p\rightarrow U^{\dag}D_p U, \quad F_p\rightarrow U^{\dag}F_p U.
\ee
If $U^{\dag} U$ is not $1$, the derivation still goes through,
and it means that the $U$ transforms
sections on the bundle $\Eps$ to the bundle $p\Eps$.
By $U^{\dag}$ one can transform sections on $p\Eps$ back to $\Eps$.

For the special case where $U$ is given by the unit $N\times N$ matrix
times $u^k$ with $u$ defined in (\ref{u}),
it transforms a bundle of constant dimension to
another bundle of dimension zero at $|n\ra$ for all $n<k$,
and dimension $N$ everywhere else.
This can be viewed as a motivation why one should not
exclude bundles of non-constant dimension.
However, there are also bundles of non-constant dimension
which can not be transformed to those with constant dimension.
The $U(1)$ instanton discussed below may be such an example.
It can be interpreted as a D0-brane on a D4-brane
in a constant B field background.
Since there are no smooth projective functions
on a connected classical space,
these bundles can not have a smooth classical limit.
For the case of $U(1)$ instanton it is suggested that
its classical interpretation via the Seiberg-Witten map \cite{SW}
is a bundle on a K\"{a}hler manifold which is
a blowup of $\C^2$ at a finite number of points \cite{BN}.
It would be interesting to see whether such a correspondence
can be established on general grounds,
not only for anti-self dual configurations.

For a matter field $\Phi$ in the adjoint representation,
$\Phi\rightarrow U^{\dag}\Phi U$.
For any state $|n\ra$,
it is always possible to find $U$ such that after
the transformation by $U$ the expectation value of
$\Phi$ at $|n\ra$ is zero.
This means that one can choose a gauge such that
the field is trivial within any given radius around the origin.
This may be a hint of holography in noncommutative gauge theories.

\section{ADHM Construction} \label{ADHM}

For completeness we review the ADHM construction
of instantons on noncommutative space \cite{NS} in this section.
Explicit expressions for some instanton configurations
are given in \cite{NS,Fur,KLY}.

Instantons are defined as anti-self dual configurations of the gauge field.
We consider the case where the spacetime coordinates
are two commuting copies of (\ref{zzb}):
\be \label{z4}
[\zh_A, \zbh_B]=\d_{AB}\th/2,\quad [\zh_A, \zh_B]=0,
\quad A,B=1,2.
\ee
This can always be achieved by linear transformations on
the coordinates if $\th$ is of maximal rank.

The ADHM construction is a prescription for finding
the conditions (\ref{xipsi}) defining the projective module.
For the $U(N)$ instanton solution of charge $k$,
the condition (\ref{xipsi}) is of the form
\be \label{Dzpsi}
D_z^{\dag}\psi=0,
\ee
where
\be
D_z=\left(\tau_z^{\dag}, \s_z\right),
\ee
with
\be
\tau_z=(B_0-\zbh_0, -B_1+\zbh_1, I),
\quad \s_z^{\dag}=(B^{\dag}_1-\zh_1, B^{\dag}_0-\zh_0, J^{\dag}).
\ee
$D_z$ plays the role of $\xi$ in sec.\ref{proj-mod}.
In order to give instanton solutions,
the constant matrices $B_0$, $B_1$, $I$ and $J$ should be solutions of
\bea
&[B_0, B_0^{\dag}]+[B_1, B_1^{\dag}]+II^{\dag}-J^{\dag}J=\th, \label{BB1}\\
&[B_0, B_1] + IJ=0, \label{BB2}
\eea
where $B_0, B_1$ are $k\times k$ matrices and
$I, J^{\dag}$ are $k\times N$ matrices.
We can write $\psi$ as
\be
\psi=\left(\begin{array}{c} \psi_0 \\ \psi_1 \\ \xi \end{array}\right),
\ee
where $\psi_0, \psi_1$ are $k$-vectors and $\xi$ is an $N$-vector.

Eq.(\ref{Dzpsi}) imposes $2k$ constraints on $\psi$,
which is a $(2k+N)$-vector,
so there are $N$ independent solutions for $\psi$,
and the projective matrix defined by solutions of $\psi$ is
a $(2k+N)\times(2k+N)$ matrix of rank $N$.
Hence we can think of the ADHM construction as a recipe
to define the $U(N)$ instanton bundle as a rank $N$ projection
from a trivial bundle of dimension $(2k+N)$.

In the case of $U(1)$ instantons,
it is impossible to find $\psi$ such that
both (\ref{compl}) and (\ref{ortho}) hold.
In order to use (\ref{FP3}), which guarantees that
the solution is anti-self dual,
one has to choose $\psi$ which satisfies (\ref{compl}-\ref{proj}).

\subsection{$U(1)$ Instanton Solution} \label{instanton}

Following \cite{NS}, we consider the $U(1)$ instanton solution
for the SHO Hilbert space.

For $N=k=1$, there is no regular classical instanton solution.
On the noncommutative space (\ref{z4}),
we can first choose $B_0=B_1=0$ by translation,
and a solution of (\ref{BB1}), (\ref{BB2}) is
\be
I=\sqrt{\th}, \quad J=0.
\ee

From (\ref{Dzpsi}), one can easily find a solution of this form
\be \label{psi01}
\psi_0=\zh_0 f, \quad \psi_1=-\zh_1 f, \quad
\xi=\frac{1}{\sqrt{\th}}(\zbh\zh)f,
\ee
where the function $f$ is arbitrary, and
\be
\zbh\zh\equiv \zbh_0\zh_0+\zbh_1\zh_1.
\ee
The operator $\zbh\zh$ has eigenvalue $\frac{\th}{2}(n_0+n_1)$
on the state $|n_0, n_1\ra$.

In order to use $\psi$ to define the projection $P$ (\ref{P}),
naively the function $f$ should be determined by (\ref{ortho})
\bea
\psi^{\dag}\psi&=& 1 \nn \\
               &=& f^{\dag}[(\zbh\zh+\th)\zbh\zh/\th]f. \label{one}
\eea
In \cite{NS} a formal solution of $f$ is given as
\be
f=f_0\equiv\left( (\zbh\zh+\th)\zbh\zh/\th \right)^{-1/2}.
\ee
This expression is in fact ill defined since it diverges
on the state $|0,0\ra$.
However, $\psi$ is well defined according to (\ref{psi01})
if the factors of $\zh_0, \zh_1$ are ordered to the right of $f$
so that the state $|0,0\ra$ is annihilated before
it causes any trouble.
It is pointed out in \cite{Fur} that this projective module
is the kind that is related to a projection $p$ in $\A$.
It is
\be \label{psiP}
\psi^{\dag}\psi=p\equiv 1-|0,0\ra\la 0,0|
\ee
on the SHO Hilbert space, and (\ref{one}) does not hold.
Another way to interpret this is that
one needs to make a further projection by $p$
to get a new projective module on which
(\ref{psiP}) is equivalent to (\ref{one}).
As we have mentioned in sec. \ref{further},
the rank of the gauge group is not a constant on the whole space.
It has rank zero at $|0,0\ra$ and rank one everywhere else.
This special property may be related to its classical interpretation
as a bundle on a blowup of $\R^4$ \cite{BN}
via the Seiberg-Witten map.

As a generalization of (\ref{u}), let
\be
u_A = (\zbh_A\zh_A+\th/2)^{-1/2}\zh_A, \quad A=0,1,
\ee
which are well defined operators satisfying
\be
u_A u_A^{\dag}=1.
\ee
(Note here that the index $A$ is not summed over.)
But $u_A^{\dag} u_A$ is not equal to $1$.
The $\th\rightarrow 0$ limit of $u_A$ is the phase factor of $z_A$,
which is ill-defined on $\C$.
When we try to solve (\ref{one}),
we can also choose $f$ to be
\be
f=f_0 U, \quad U=u_0^m u_1^n
\ee
for non-negative integers $m, n$.
This results in a change of $\psi$ by $\psi\rightarrow \psi U$,
and is thus an example of the situation discussed in sec.\ref{further}.

On the other hand,
if we choose $U$ to be given by products of $u_A^{\dag}$,
eq.(\ref{ortho}) is satisfied,
but the condition (\ref{compl}) is not satisfied for $\xi=D_z$.
($\xi$ is related to $\phi$ by a linear transformation.)
In order to use (\ref{FP3}),
one has to adjoin new vectors to $\phi$,
making it an $N_0\times (N_0-N+1)$ matrix,
and the result is not anti-self dual anymore.

\section*{Acknowledgment}

I would like to thank Keun-Young Kim, Bum-Hoon Lee,
Kimyeong Lee, Miao Li, Hyun Seok Yang and Piljin Yi
for discussions.
I also thank APCTP and KIAS where part of this work is done
for hospitality.
This work is supported in part by
the National Science Council, Taiwan, R.O.C.
and the Center for Theoretical Physics at
National Taiwan University.

\vskip .8cm

\end{document}